\documentstyle[11pt,mrs2001,epsfig]{article}
\def\msun{\mbox{ M}_\odot}
\def\mag{\mbox{ mag}}
\def\kms{\mbox{ km s$^{-1}$}}

\def\kpc{\mbox{ kpc}}

\begin{document}

\title{The Dark Galaxy Problem and the Effects of Substructure on
Gravitational Lenses}

\author{R. Benton Metcalf}

\affil{Institute of Astronomy \\
Madingley Road \\
Cambridge CB3 0HA \\
England}

\begin{abstract}
I argue that the cold dark matter (CDM) model requires that even within
a few kpc of the center of a galactic halo a significant fraction
(greater than a few percent) of the surface density is contained in
substructures with masses $\ga 10^3 \msun$.  These structures should be
light enough to avoid dynamical friction and dense enough to avoid tidal
disruption.  I then show using the results of numerical simulations that
this substructure will significantly alter the flux ratios of multiply
imaged quasars (QSOs) without changing the image positions.  The degree
to which this occurs will depend on the angular size of the QSO and thus
the wavelength of the observations. 
\end{abstract}

\section{Introduction}

The $\Lambda$CDM model of structure formation has so far been enormously 
successful in explaining the large scale structure of the universe and
the anisotropies in the cosmic microwave background (CMB).  However the
model appears to face some difficulties explaining observations on galactic
and smaller scales.  A number of such problems have been discussed in the
literature \cite[for example]{MQGSL,KKBP,BS01,SK}.  For the present discussion the most
relevant of these problems is the observation that CDM simulations of
the local group of galaxies predict an order of magnitude more dwarf
galaxy halos with masses greater than $\sim 10^8\,\msun$ than are observed
\cite{1999ApJ...524L..19M,1999ApJ...522...82K,Mateo}.  This could be a
sign that there is something fundamentally wrong with the CDM model
\cite{Bode01,2001ApJ...547..574D,2000PRL.Kamionkowski}.  Alternatively, the small DM clumps could exist, but
not contain observable dwarf galaxies.  This situation can easily,
perhaps inevitably, come about through the action of feedback processes
in the early universe.  For example, photoionization can prevent gas
from cooling and thus inhibit star formation in small halos
\cite{2000ApJ...539..517B}.
In section~\ref{sec:abundance-small-mass}, I show that the
overabundance of DM clumps with respect to visible galactic satellites is
likely to extend down to smaller masses and larger fractions of the halo
mass than have thus far been accessible to numerical simulations.  

These nearly pure dark matter structures have largely been considered
undetectable (see \cite{TMR01} for a review of possible methods).
Acting as individual gravitational lenses clumps with velocity
dispersions of $10\kms$ (corresponding to a mass of order $10^8\msun$)
rarely create multiple images and when they do the image separations are
too small to be resolved (milli--arcseconds) and are not variable on an
observable time scale.  However, as shown in \cite{MM01}, if the CDM
model is correct and these substructures exist within the lenses
responsible for multiply imaged QSOs they will have a dramatic effect on
the image magnifications.  This is the subject of
section~\ref{sec:comp-grav-lens}.

\section{The abundance of small mass substructure}
\label{sec:abundance-small-mass}

\begin{figure}
\plotone{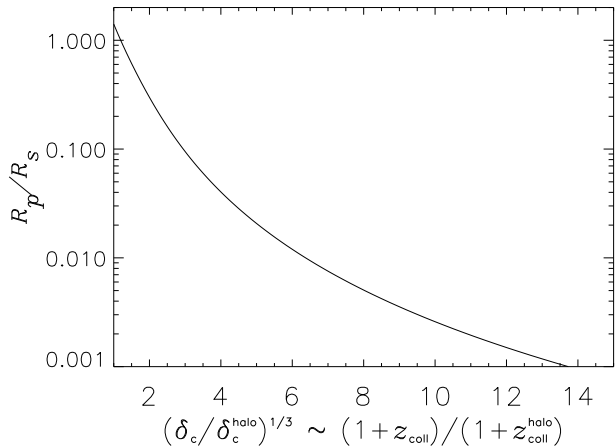}{3.5in}
\caption{\footnotesize The penetration depth of a
subclump as a function of the collapse redshift of the subclump,
$z_{coll}$, and the collapse redshift of the host halo,
$z^{halo}_{coll}$.  The scale size of the host halo is $R_s$ (about
20~kpc for the Milky Way).}
\label{fig:depth}
\end{figure}

In the CDM model galaxy halos are built through the merger of
smaller halos.  The larger of these subclumps ($\la 10^9\msun$) sink
into the center of the halo through the action of dynamical friction
where they are destroyed by tidal forces.  The orbits of smaller clumps
decay much less quickly, but they will suffer tidal stripping and
collisions.  The Nbody simulations of Klypin 
{\it et al.} \cite{1999ApJ...522...82K} and Moore {\it et al.}
\cite{1999ApJ...524L..19M} have shown that substructures of masses $\ga
10^7\msun$ do survive in significant numbers in galactic halos.  Within
the virial radius of a galactic halo roughly 10\% to 15\% of the mass is
contained in such subclumps.  Here I consider how many smaller mass
subclumps will survive.

Navarro, Frenk \& White \cite{1997ApJ...490..493N} (NFW) found that CDM
clumps in their simulations have universal spherically averaged profiles of
the form 
\begin{equation}
\rho(r)=\frac{\rho_{crit}\delta_s}{(r/r_s)(1+r/r_s)^2}.
\end{equation}
Once the cosmological parameters and the initial spectrum of density
fluctuations have been chosen the NFW profile is approximately a one
parameter family; we can consider $r_s$ to be a function of $\delta_s$.
There is a significant scatter about this relation however.  In
agreement with the spherical collapse model the central density is found
to be proportional to the average density of the universe at the time
when the clump became nonlinear, the collapse time, so $\delta_s\propto
(1+z_{coll})^3$.

Clumps orbiting within a larger halo will have mass stripped off them
until their radius reduces to approximately the tidal radius, $r_t$, defined by
\begin{equation}
r_t(\delta_s,R,\delta_s^{halo}) = R \left( \frac{m(r_t)}{M(R)} \right)^{1/3}
\left[3-\frac{\partial\ln M}{\partial\ln R}\right] ^{-1/3}
\end{equation}
where $M(R)$ is the mass of the host halo interior to $R$ and $m(r)$ is
the same for the subclump.  The superscript ``$halo$'' refers to the host halo.  It was shown in \cite{MM01} that while the
tidal radius is larger than the subclump scale length, $r_s$, tidal
stripping does not remove a significant amount of mass from the
subclumps.  With this in mind a penetration depth, $R_p$, can be defined by the
relation 
\begin{equation}
r_t(\delta_s,R_p,\delta_s^{halo})=r_s(\delta_s).
\label{eq:depth}
\end{equation}
At galactic radii $R\la R_p$ mass is transfered between the
substructures and a smoother component of DM.  Since subclumps are not
generally on circular orbits $R_p$ should be thought of as the smallest
pericenter distance a clump can have during its orbits without losing
much of its mass.  It turns out that $R_p$ can be expressed in terms of just
$\delta_s/\delta^{halo}_s = (1+z_{coll})^3/(1+z^{halo}_{coll})^3$.  The
penetration depth is plotted in figure~\ref{fig:depth}.
It can be seen there that if $(1+z_{coll})/(1+z^{halo}_{coll}) \sim 3$
or larger the subclump can remain intact down to within a few kpc from the
center of the host halo.  If we are interested in a galactic halo that
is assembled at $z^{halo}_{coll}\sim 1$ then a small mass clump that
formed at $z_{coll} \ga 6$ is likely to survive within it.

\begin{figure}
\plotone{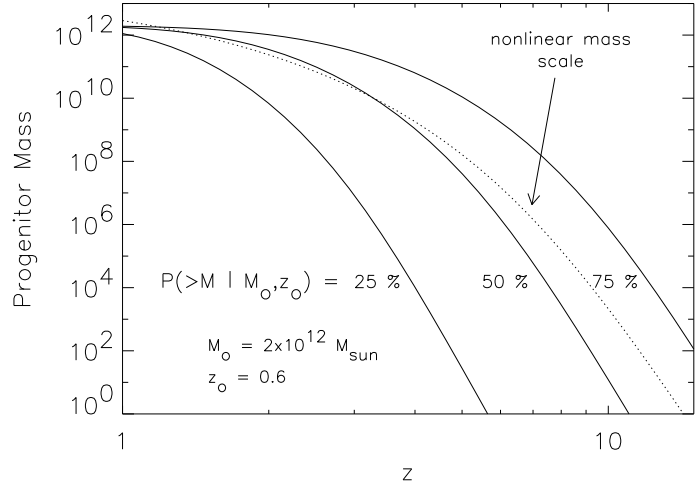}{4.in}
\caption{\footnotesize This plot concerns the
distribution of progenitors of a halo that is $2\times 10^{12}\msun$ at
$z=0.6$.  The given fraction (25\%, 50\% and 75\%) of the halo mass is
contained in clumps of masses smaller than the curve at the given
redshift.  For example, 75\% of the mass that will end up in the halo is
contained in separate clumps with masses $<10^6\msun$ at $z=10$.
Plotted as a dotted curve is the typical mass of objects undergoing
nonlinear collapse.}
\label{fig:progen}
\end{figure}

How plentiful should such subclumps be?  The merger history of a halo 
can be calculated using the extended Press-Schechter formalism
\cite{1993MNRAS.262..627L}.  Figure~\ref{fig:progen} shows what fraction
of the mass in a typical galactic halo at $z=0.6$ is 
in separate clumps of a given mass range at higher redshifts.  The
collapse of a typical galaxy halo of mass $\sim 10^{12}\msun$ becomes
nonlinear at $z\sim 1$ in the $\Omega_{\rm matter}=0.3$,
$\Omega_\Lambda=0.7$ model used here.  Fifty percent of this halo is in clumps of mass
$m\la 10^6\msun$ at $z=6$ which must have undergone collapse at some
earlier time (typically this mass scale collapses at $z\sim 7$ though in
the overdense region surrounding the future large halo the mean $z_{call}$
could be larger).  Even at $z=2$, 15\% of the final halo is in free
floating clumps of $m\la 10^7\msun$ which typically collapsed before
$z\sim 6$.  

It is difficult to imagine how all this very compact substructure could
be efficiently destroyed within the final halo.  Collisions are not
effective because they are rare and generally at too high a speed to
disrupt the clumps.  Even when (\ref{eq:depth}) is violated the lose of
mass in substructure is a steeper, but still gradual function of $R$.
In addition, the observational constraints on dark satellites of the
Milky Way are rather weak in this mass range \cite{MM01}.  As a result I
think it is reasonable to assume that at least several percent, perhaps
significantly more, of the surface density at a projected radius of
$R\sim \kpc$ is contained in substructure.  A firmer prediction will
ultimately require simulations with mass resolutions several orders of
magnitude smaller than are now available.

\section{Compound gravitational lensing}
\label{sec:comp-grav-lens}

Although the gravitational pull on a light beam from a $m\la 10^9\msun$
DM clump is very weak, if the beam also passes through a larger mass
concentration the contributions to the magnification from the two
structures will combine nonlinearly and enhancing the influence of the
smaller clump.  With the larger lens producing multiple images the
effects of the small clumps can be seem in the magnification ratios.
For this to happen the angular size of the source cannot be much
larger than that of the substructures.  QSOs are the obvious choice.

The large number of subclumps and the small, but not vanishingly small,
size of the source make numerical simulations of the lensing necessary.
Generally an image is influenced by several subclumps acting together
rather than an individual one.  As a simple example we consider a
singular isothermal sphere (SIS) for the primary lens, $M(R)\propto
R$.  Two images are formed, one farther 
from the center of the lens, image~1, and one closer, image~2.  Image~2 is
reversed in one dimension with respect to image~1 and the original
source.  The Einstein ring radius which characterizes the image
separations is $r_E = 4\pi (\sigma_{halo}/c)^2D_{\rm ls}D_{\rm l}/D_{\rm
s}$ on the lens plane.  $D_{\rm s}$, $D_{\rm l}$ and $D_{\rm ls}$ are the
angular size distances to the source, the lens and between the source
and lens.  The one--dimensional velocity dispersion of the host halo is
$\sigma_{\rm halo}$.  In a flat $\Omega_{\rm matter}=0.3$ model $r_E =
3.5~ h^{-1} \kpc$ or 0.3 arcseconds for a lens with $\sigma_{halo}=150
\kms$ at $z=1$ and a source at $z=3$.

In our simulations the subclumps are positioned randomly, but with an
average number density proportional to the total surface density of the primary
lens.  They are truncated at the tidal radius appropriate for their
internal structure and position.  This neglects the fact that subclumps
may be on elliptical orbits which have taken them closer to the
center halo in the past.  The mass spectrum of the clumps is taken to be
$dN/dm \propto m^{-2}$ in accordance with numerical simulations that
extend down to $10^7\msun$.  The lower mass cutoff in the simulations is
effectively set by the smallest mass that can have a significant effect
on the lensing,
\begin{equation}
m_c \simeq \frac{2c^3R}{\sqrt{3}G\sigma_{\rm halo}}
\left(\frac{l_{\rm s}}{4\pi D_{\rm ls}}\right)^{3/2}
\label{eq:mc}
\end{equation}
where $l_s$ is the proper size of the source.  Because of this cutoff
the source size dictates what mass scales are being probed by compound
lensing.  For the same example used before $m_c \simeq 10^4\msun (l_s
h/{\rm pc})^{3/2}$ at $R=1\kpc$.  The broad--line region of a QSO is believed
to be a few pc in size.  For a more detailed description of these numerical
simulations see \cite{MM01}.

The simulations show that the centroid positions of the images are not
change significantly by the substructure unless they happen to lay very
near a subclump of mass $\sim 10^8\msun$ or larger which is a rare
occurrence.  Because of this the image positions can still be used to
constrain the overall shape of the halo.  The resulting ``smooth model''
can then be used to predict the magnification ratios of quadruply imaged
QSOs.  This is a more difficult task than might be assumed because
smooth models that fit the positions well are usually highly degenerate,
predicting a large range of magnification ratios.  However, progress has
been made in this direction which will be reported elsewhere \cite{MZ01,2001MNRAS.320..401Z}.

\begin{figure}
\vspace{-1.25in}
\plotone{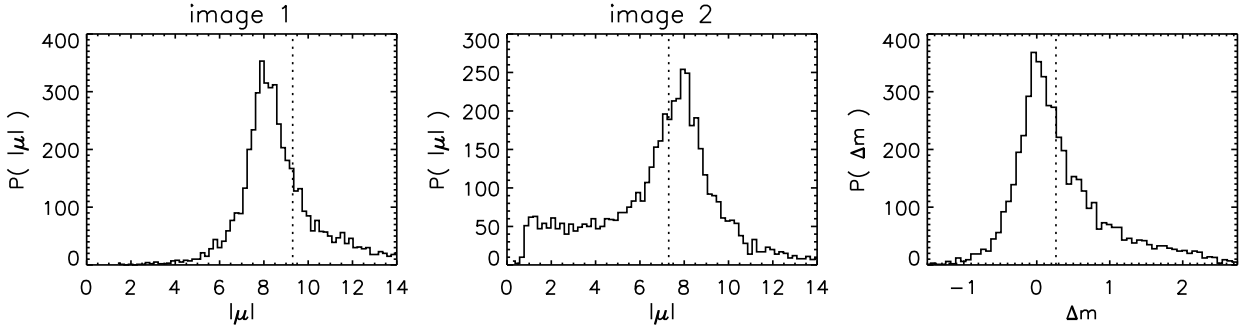}{6.6in}
\vspace{-1.6in}
\caption{\footnotesize The two panels on the left are the magnification
distributions for the two images of a QSO at $z=3$ lensed by a singular
isothermal sphere at $z=1$.  On the right is the distribution of magnification ratios
expressed in magnitudes, $\Delta m \equiv 2.5\log(|\mu_1|/|\mu_2|)$.  The
subclumps have masses of $10^4\msun < m < 10^8\msun$ and constitute 5\% of
the halo surface density.  The vertical lines show the values expected
for the same lens without substructure.  The source is at a projected
distance of 0.12 Einstein ring radii form the center of the lens.  The
source is 10~pc in radius.  5,000 realizations are shown.}
\label{fig:mag_dist} 
\end{figure}

Sample magnification and magnification ratio probability distributions
are shown in figure~\ref{fig:mag_dist}.  These are produced by creating
random realizations of the subclump positions and masses.  Images~1 and
2 are affected differently by the substructure.  This is a result of the one
dimensional parity flip of image~2.  The most likely $\mu$ is biased in
opposite directions for the two cases which makes the magnification
ratio biased low; that is closer to equal magnification.  Negative
values of $\Delta m$ are cases where the order of brightness is the
opposite of what is expected from the smooth model.  The spread in
$\Delta m$ is quite large compared to typical 
observational errors, $\sim 0.01\mag$.  In this example $\Delta m$ is more
than 0.2 mag from the smooth model in 74\% of the cases.  Note that
subclumps contain only 5\% of the surface density in this example.  With
sufficiently secure smooth models the absence of such substructure would
be clear in the existing sample of quadruply imaged QSOs if it
were not for microlensing by ordinary stars.  

There are other approaches to detecting compound lensing that could
also be pursued.  One is to look at the
statistics of QSO magnification ratios in general.  Substructure will
create an overabundance of near equally magnified images.  Substructures
will also deform multiply imaged radio jets on milli-arcsecond scales.
The images can be comparing for signs of small scale distortions that are
inconsistent with a smooth lens.  These approaches are discussed further
in \cite{MM01}.

\section{Discussion}

The continuum emission region of a QSO in visible wavelengths is very
small ($\la 1,000$ AU) and as a result it is susceptible to microlensing
by ordinary stars.  It has been established that microlensing does cause
changes in the image brightness of Q2237+0305 as large as a magnitude
\cite{2000ApJ...529...88W}.  This could potentially wash out any signal
from DM substructures.  Fortunately the emission in radio and in atomic
and molecular lines are believed to come from a larger region.  Variability in
the QSO could also be a complication.  The time delays between images
range from a fraction of a day to months.  As long as the source is more
then a light-month in size it should not vary on a timescale that would
interfere with these investigations.  These considerations demonstrate the
need for magnification ratios measured in multiple frequencies.  With
coming data we may soon have a new and unique test of the CDM model.

\acknowledgements{I would like to thank my collaborator on much of this
work Piero Madau.}

\vfill

\begin{thebibliography}{}{

\bibitem{Bode01} Bode, P., Ostriker, J. P., \& Turok, N. 2001,
\apj, 556, 93

\bibitem{BS01} Borriello, A. \& Salucci, P., 2001,
\mnras, 323, 285

\bibitem{2000ApJ...539..517B} {Bullock}, J.~S., {Kravtsov}, A.~V., \&
{Weinberg}, D.~H. 2000, \apj, 539, 517; Somerville, R. S., 2001,
submitted to \apj (astro-ph/0107507)

\bibitem{2001ApJ...547..574D} Dav{\' e}, R., Spergel, D.~N., Steinhardt,
P.~J. \& Wandelt}, B.~D., 2001, \apj, 547, 574

\bibitem{2000PRL.Kamionkowski} Kamionkowski, M. \& Liddle, A.,2000,
\prl, 84, 4525

\bibitem{KKBP} Klypin, A., Kravtsov, A.~V., Bullock, J. S., \& Primack, J. R. 2001, ApJ, 554, 903

\bibitem{1999ApJ...522...82K} {Klypin}, A., {Kravtsov}, A.~V., {Valenzuela}, O., \& {Prada}, F.  1999, \apj, 522, 82

\bibitem{1993MNRAS.262..627L} Lacey, C. \& Cole, S., 1993, \mnras, 262, 627

\bibitem{Mateo} Mateo, M. 1998, \araa, 36, 435

\bibitem{MM01} Metcalf, R.B. \& Madau, P., 2001, \apj ~in press, (astro-ph/0108224)

\bibitem{MZ01} Metcalf, R.B. \& Zhao, H., 2001, in preparation.

\bibitem{1999ApJ...524L..19M}{Moore}, B., {Ghigna}, S., {Governato}, F., {Lake}, G., {Quinn}, T., {Stadel},  J., \& {Tozzi}, P. 1999, \apjl, 524, L19

\bibitem{MQGSL} {Moore}, B., {Quinn}, T., {Governato}, F.,
{Stadel}, J., \& {Lake}, G. 1999, \mnras, 310, 1147;  

\bibitem{1997ApJ...490..493N} Navarro, J.~F., Frenk, C.~S., \& White,
S. D.~M. 1997, \apj, 490, 493

\bibitem{SK} Sellwood, J., \& Kosowsky, A. 2001, in Gas and Galaxy Evolution 
(ASP Conf. Series), in press (astro-ph/0009074) 

\bibitem{TMR01} Trentham, N., Moller, O. \& Remirez-Ruiz, E., 2001,
\mnras ~in press (astro-ph/0010545)

\bibitem{2000ApJ...529...88W} Wo{\' z}niak, P.~R., Alard, C., Udalski,
A., Szyma{\' n}ski, M., {Kubiak}, M., Pietrzy{\' n}ski, G. \& Zebru{\'
n}, K., 2000, \apj, 529, 88

\bibitem{2001MNRAS.320..401Z} Zhao, H. \& Pronk, D., 2001, \mnras, 320, 402
\end{thebibliography}
\end{document}